\documentclass[man, floatsintext, 12pt]{apa7}
\usepackage[utf8]{inputenc}
\usepackage{tabularx}
\usepackage{longtable}
\usepackage{booktabs}
\usepackage{graphicx}
\usepackage{amssymb}
\usepackage{float}
\usepackage{url}
\usepackage{natbib}
\usepackage[doublespacing]{setspace}
\usepackage{tikz}
\usetikzlibrary{shapes.geometric, arrows}

\title{Optimal item calibration in the context of the Swedish Scholastic Aptitude Test}
\shorttitle{Optimal item calibration - SweSAT}
\title{Optimal item calibration in the context of the Swedish Scholastic Aptitude Test}
\shorttitle{Optimal item calibration SweSAT}
\authorsnames{Jonas Bjermo$^{1,2}$,  Ellinor Fackle Fornius$^1$, Frank Miller$^{1,2}$}

\authorsaffiliations{{$^1$Department of Statistics, Stockholm University, Sweden}\\
{
$^2$Department of Computer and Information Science, Link\"oping University, 
Sweden
}}

\authornote{\addORCIDlink{Jonas Bjermo}{0000-0001-7552-8983}\\
\addORCIDlink{Ellinor Fackle Fornius}{0000-0003-0528-0083}\\
\addORCIDlink{Frank Miller}{0000-0003-4161-7851} \\
Correspondence concerning this article should be adressed to Ellinor Fackle Fornius, Department of Statistics,
Albanovägen 12, 114 19 Stockholm, Sweden. Email: \href{mailto:ellinor.fackle-fornius@stat.su.se}{ellinor.fackle-fornius@stat.su.se}}

\abstract{Large scale achievement tests require the existence of item banks with items for use in future tests.
Before an item is included into the bank, its characteristics need to be estimated. The process of estimating the item characteristics is called item calibration.
For the quality of the future achievement tests, it is important to perform this calibration well and it is desirable to estimate the item characteristics as efficiently as possible. Methods of optimal design have been developed to allocate calibration items to examinees with the most suited ability.
Theoretical evidence shows advantages with using ability-dependent allocation of calibration items. However, it is not clear whether these theoretical results hold also in a real testing situation. In this paper, we  investigate the performance of an optimal ability-dependent allocation in the context of the Swedish Scholastic Aptitude Test (SweSAT) and quantify the gain from using the optimal allocation. On average over all items, we see an improved precision of calibration. 
While this average improvement is moderate, we are able to identify for what kind of items the method works well. 
This enables targeting specific item types for optimal calibration. 
We also discuss possibilities for improvements of the method.
}

\begin{document}

\maketitle

\setcounter{secnumdepth}{3}

\section{Introduction}\label{introduction}

Item calibration is the process of estimating the characteristics of new test items. The goal of item calibration is to develop a bank of items with precisely estimated item parameters ready for use in operational tests. It is of great importance that the item parameters are estimated as precisely as possible, since it directly affects the accuracy and estimated standard errors of the latent ability estimates \citep{cheng2010impact}.
As \cite{ali2014item} point out, well calibrated items are particularly important for computerized adaptive testing (CAT), when test items are assigned to the examinees adaptively based on gradually updated estimates of their latent ability. In this setting it is assumed that the item parameters have been estimated with enough precision to be treated as the true ones \citep{Linden2000}. 

Methods of optimal experimental design can be applied in this context to determine which examinees to select from a population such that the item parameters are estimated as efficiently as possible. The precision of the item parameter estimates depends on the ability levels of the examinees that respond to the item. The design problem here is the problem of selecting a sample of examinees with the most suitable abilities, that is to find an optimal sampling design \citep{buyske2005optimal, berger1991efficiency}. In addition to being more cost-effective, using an optimal sampling design means that the item assignment will be targeted to examinees in a better way in terms of item difficulty compared to when assigning items randomly. Assigning items with a fitting difficulty level reduces the burden of the examinee taking the test and the risk of identification of the calibration items. 

Which sampling design strategy that is feasible depends on whether the calibration is conducted in an “online” or “offline” setting and whether the calibration items can be administered adaptively. In an online setting, calibration items are integrated in a test together with the operational items \citep{Stocking1988}. This allows items to be assigned to the examinees in an adaptive way, where the rule of the assignment is determined by an optimal design.  Alternatively, the calibration can be done in a separate test, consisting of calibration items only, a so called offline calibration \citep{He2020}.

The online calibration setting is well suited for adaptive optimal design schemes adjusted to sequentially arriving examinees, sequential calibration methods are implemented in different ways by \cite{Linden2000}, \cite{chang2010online}, \cite{ali2014item}, \cite{Lu2014} and \cite{Linden2015}. \cite{zheng2017comparison} compared five methods for calibration in an online sequential setup. While the considered methods that are based on an optimality criterion in theory should improve parameter precision, they observed in their simulation study that a random design in many situations achieved a comparable performance. 

However, not all testing situations have sequentially arriving examinees. Instead, the test is conducted by a large number of examinees taking the test simultaneously in parallel. Such a parallel settings is common for large-scale achievement tests, e.g. for the Programme for International Student Assessment \citep[PISA;][]{PISA} and for the Swedish Scholastic Aptitude Test \citep[SweSAT;][]{UMU} considered in this paper.  For item calibration under such parallel test settings other optimal design methods are needed. The method for calibrating items based on an optimal design algorithm by \cite{ULHASSAN2021107177} is suitable for calibration of items in a parallel setting. The method is designed for allocating a large group of examinees in parallel to items based on examinee ability. The algorithm utilizes a so called optimal restricted design. The optimal restricted design creates a set of non-overlapping ability intervals that dictates which examinee should calibrate which item. Once the ability of an examinee is estimated, it is known which ability interval the examinee belongs to, and the designated item can be allocated to the examinee.

The computations involved in the optimal design algorithm are based on some approximations which means the demonstrated theoretical efficiency gains are not guaranteed to hold in practice. The optimal design therefore needs to be evaluated in a real testing situation. Moreover, \cite{zheng2017comparison} demonstrate that a random assignment can give comparable results as optimal designs when they evaluated several different test conditions.

In this study, we evaluate the optimal allocation strategy proposed in \cite{ULHASSAN2021107177} and compare the results to the random allocation strategy. We use real data from the Swedish Scholastic Aptitude Test (SweSAT) to conduct an empirically-based simulation study,  designed to replicate the real calibration setting of SweSAT. 
Moreover, we consider a set of different simulation scenarios with varying degree of assumptions made. In this way, we are able to separate the effect of each assumption. The responses in the test round of 2018 are used to estimate item parameters that are utilized in the optimal design algorithm to decide the optimal allocation on the basis of estimated examinee ability. The calibration procedure of SweSAT is that each examinee calibrates the same fixed number of items, in contrast to alternative schemes where the items may be given to the examinees until the standard errors of the estimates meet some criteria, or by letting a fixed number of examinees calibrate each item \citep{Linden2015,Ren2017}.

The abilities used for assigning the examinees to items are estimated from an operational test. A straightforward way is to treat the ability estimates, used for item assignment, as the true values of the abilities as described by  \cite{Chen2016} and \cite{Stocking1988}. These abilities are treated as fixed and used for estimating the item parameters. This produces item parameter estimates of the calibration items to appear on the same scale as the operational items. 


The item response theory (IRT) models that are being used to describe the relationship between the items and latent abilities depend on the model parameters in a nonlinear way. This means that the optimal allocation design is dependent on the unknown item parameters that are supposed to be estimated, see for example \cite{Atkinsson}. The item parameters, therefore, need to be either pre-estimated or given some values by expert guessing; for references justifying the latter see \cite{berger2019efficiency}. We explore the effect of pre-estimation in the simulation study by comparing the precision of estimates when the optimal design is derived assuming the true parameter values known to when it is based on pre-estimated values. The SweSAT consists of multiple choice items and we fit the three-parameter logistic (3PL) IRT model.
We note that \cite{berger2019efficiency} investigated and compared by a simulation study calibration designs which can be used when examinees take the test simultaneously in parallel. Their designs are not based on optimality criteria and they assume the Rasch model for the calibration items. They investigate the impact of item-parameter pre-estimation. 

The article is organized as follows. We start with describing the 3PL model and the proposed optimal allocation method and give some background about the SweSAT. Next, the details about the setup of the empirically-based simulation study, and its different scenarios, are specified along with the definitions of the measures and evaluation metrics used to compare the optimal and random designs. Then the results from the simulation study are presented, and the paper ends with a discussion about the results. 
\bigskip

\subsection{The 3PL Model}\label{3PLmodel}

The test items considered here are all multiple-choice questions that have dichotomous outcomes; a response is either correct or incorrect. Therefore, a 3PL model \citep[see e.g.\ ][]{lord1980applications} is used to model how the probability of a correct response depends on examinee ability. The 3PL model has 3 item parameters $\beta = \left(a, b, c\right)^\top$; the $a$-parameter is related to item discrimination, $b$ reflects item difficulty and $c$ is the so-called 'guessing' parameter.  For the 3PL model the probability that an examinee with ability $\theta_j$ correctly responds to item $i$  with parameters $\beta_i = \left(a_i, b_i, c_i\right)^\top$, is

\begin{equation} \label{eq1}
  p_{i}\left(\theta_j\right)=p_{i}\left(\theta_j|a_i,b_i,c_i\right)=c_i+\frac{1-c_i}{1+e^{-a_i\left(\theta_j-b_i \right)}}.
\end{equation}

It can be noted that the 3PL model is a weighted version of the 2PL model, being weighted with the guessing parameter $c_i$. The 2PL model is a generalized linear model with logit link $\eta_i(\theta) = \log\left(\frac{p_i(\theta)}{1-p_i(\theta)}\right)$ where $p_i$ is the probability defined in Equation \ref{eq1}. The link is differentiable in $\beta_i$.

\begin{figure}[H]
\centering
\includegraphics[scale=0.6]{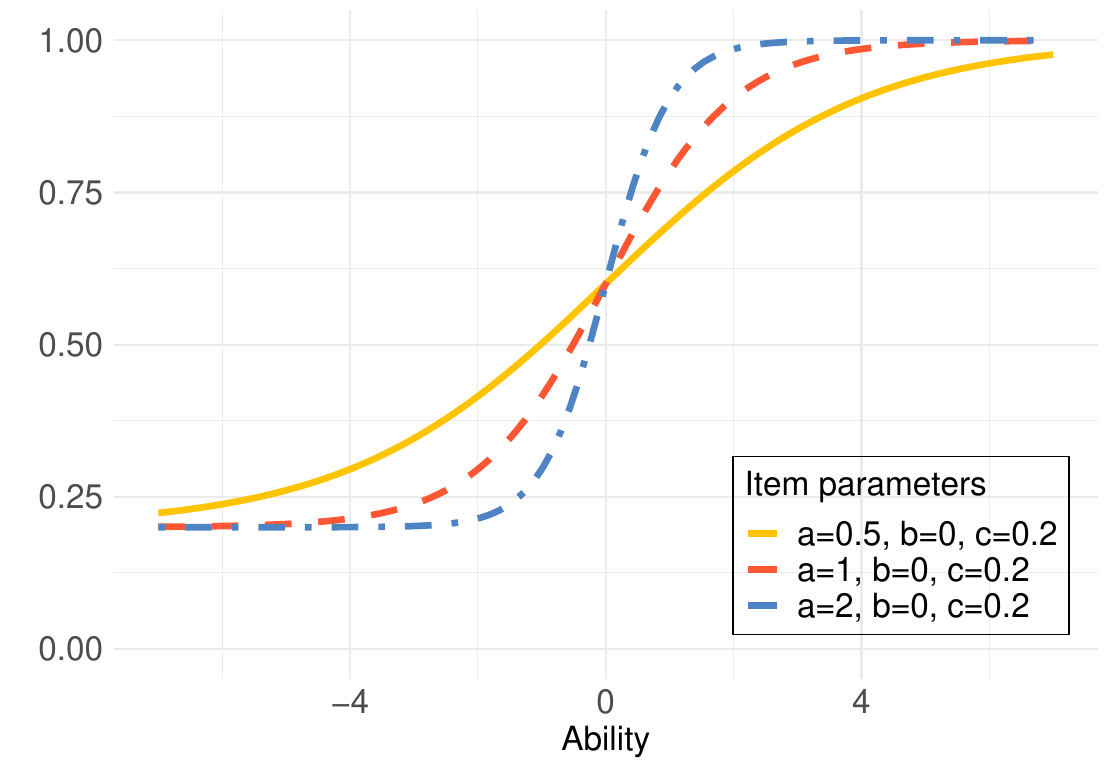}
\caption{Item response curves with varying discrimination levels}
\label{fig:3IRT}
\end{figure}

The graph of the probability to correctly respond to an item as a function of examinee ability $\theta$ illustrates the item characteristics and is often referred to as the \textit{Item Characteristic Curve (ICC)} or item response curve. Figure \ref{fig:3IRT} displays three item response curves with different levels of item discrimination. 

\subsection{Optimal design allocation}\label{Optimaldesign}

An optimal design of an experiment determines the experimental conditions in a way that the estimation is optimal with respect to some criterion, see for example \cite{Silvey}. Methods of optimal experimental design are used here to derive an 
optimal allocation strategy for item calibration. The optimal design allocation provides rules for matching calibration items to examinees based on examinee ability, such that the precision of item parameter estimates is optimized. 

Using an optimal design allocation, the item parameters in the 3PL model will be estimated with better precision compared to a non-optimal design, at least in theory. Since the item parameter estimates are correlated \citep{Wingersky1984}, we do not focus on the precision of the three model parameters separately. Rather we use here the so-called D-optimality criterion \citep{Atkinsson} which takes the correlation of the parameters into account. This criterion minimizes the determinant of the inverse information matrix (asymptotically equivalent to the covariance matrix) of the item parameter estimators. This determinant is proportional to the volume of a confidence ellipsoid for the three model parameters.

A standard (unrestricted) optimal design specifies a number of ability levels that would be optimal to sample from, a design where equal proportions of the examinees are divided between two specific $\theta$ points would be an optimal design under the 2PL model. Such an unrestricted design is feasible if there are no restrictions on the availability of examinees with certain abilities.  However, in a real test situation, it is not realistic to be able to choose an examinee with the exact ability needed. Instead, \cite{Hassan2019} proposed a restricted design that is more reasonable to attain in practice. In the restricted design, ability intervals are instead specified, as opposed to ability points for the unrestricted optimal design. 
While \cite{Hassan2019} have exemplified the restricted design approach for the 2PL model, it is valid even for other IRT models including the 3PL model; the latter model has been considered by \cite{ULHASSAN2021107177}.

Let $g$ be a continuous density on $\Theta = {\rm I\!R}$ which describes the abilities of the examinees; we assume in this article that the examinees have standard normal distributed abilities and $g$ is the $N(0,1)$-density. A restricted design is described by sub-densities $h_1,h_2,\dots ,h_n \geq 0$ for each item in the test, where $\sum_{i=1}^n h_i(\theta) = g(\theta)$ for all $\theta \in \Theta$. Each $h_i$ represents the part of the examinee population calibrating item $i$. 

The standardized information matrix of the item parameters $\beta = (\beta_1,\dots,\beta_n)$ is $M(h) = \mbox{diag}(M_1(h_1),..., M_n(h_n))$ with 
\begin{equation} \label{eq2} M_i(h_i) = \int_{\Theta}p_i(\theta)(1-p_i(\theta))\left(\frac{\partial\eta_i(\theta)}{\partial \beta_i} \right)\left(\frac{\partial\eta_i(\theta)}{\partial \beta_i} \right)^T h_i(\theta)d\theta. \end{equation}
Here, the density $h$ summarizing the sub-densities $h_i$ describes the allocation rule saying which item $1,\dots,n$ should be calibrated by examinees with a specific ability $\theta$ (formally, it is defined as $h(\theta, i)=h_i(\theta)$ and is a density on the product space $\Theta \times \{1, \dots, n\}$). Also, $\partial\eta_i(\theta)/\partial \beta_i$ is the derivative of the logit link described in Section \ref{3PLmodel} with respect to $\beta_i$.

To obtain an optimal design, an appropriate convex function $\Psi$ of $M(h)$ needs to be optimized; a design $h^*$ is $\Psi$-optimal if $h^* = \mbox{arg min}_h \Psi(M(h))$.

The considered model is not linear in this case; it means that $M(h)$ is dependent on the item parameters $\beta_i$. Therefore, some initial values must be assigned to the item parameters $\beta_i$. Initial values can be obtained by a guess from an expert or some pre-estimation using a small sample of examinees. The optimal design $h^*$ is said to be a locally optimal restricted design \citep{Atkinsson,Hassan2019}. A sample of 30 examinees was enough for pre-estimation of item parameters in the situation considered by \cite{he2020optimal}.

As mentioned above, the D-optimality criterion is used which is computed by minimizing
\begin{equation}\label{eq3}
\Psi\{M(h)\} = -\log|{M(h)}| = -\left(\sum_{i=1}^n \log(|M_i(h_i)|)\right).\end{equation}

\begin{figure}[!ht]
\centering
  \includegraphics[width=\linewidth]{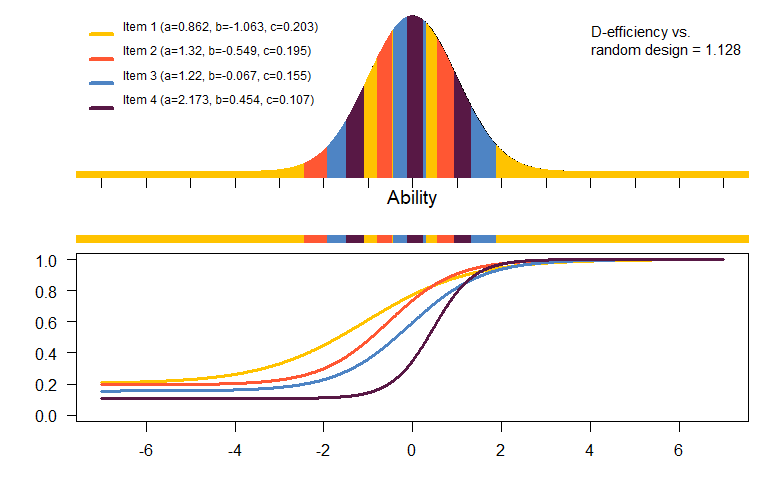}
  \caption{Calibration of 4 items under the 3PL model. Upper panel: The different colors of the normal distribution depict the ability level intervals of the locally D-optimal design for these items. Lower panel: Assumed item response curves for the 4 calibration items.}
  \label{fig:optalgplot}
\end{figure}

\cite{Hassan2019} derived a new equivalence theorem for calibration of multiple items which 
is able to identify if a design is D-optimal. To find an optimal design, an exchange algorithm  can be used \citep{ULHASSAN2021107177} which is implemented in the \texttt{R}-package \texttt{optical} \citep{optical2023}. An assumption in the algorithm is that the calibration items are arranged in blocks and that each examinee can calibrate one item in each block. If e.g. four items are given, the exchange algorithm will find the optimal design for those specific items. The design defines which item an examinee with a certain ability will be given in a calibration situation, see Figure \ref{fig:optalgplot} for an example of a produced design. E.g., this design recommends that an examinee should receive Item 3 who has, based on the operational items, an ability in one of the four ability intervals: $[-1.92, -1.49], [-0.43, -0.13], [0.23, 0.31], [1.32, 1.88]$. For this example, the theoretical D-efficiency of this D-optimal design compared to a design which randomly allocates the items to the examinees is 1.128. This means that the random design needs 12.8\% more examinees to attain similar information about the parameters compared to the optimal design, see \cite{ULHASSAN2021107177}.

\subsection{The Swedish Scholastic Aptitude Test (SweSAT)}\label{swesat}
SweSAT is a standardized test used for admission to higher education in Sweden \citep{UMU}. The test is given twice a year to around 40 000 examinees. It consists of two main sections - a quantitative and a verbal section. The test is a paper and pencil-based test with a total of five parts; two quantitative parts and two verbal parts, and one try-out/calibration part used for testing the performance of new items. Every part consists of 40 items, all equivalent across all test locations except for the try-out parts which could differ between test locations. The total score is put on a scale from $0$ to $2$ and is equated between tests to be comparable. 

In this study, we use response data from the quantitative part of the second test round of 2018 and estimate the item parameters which are used as a starting point in our simulation study. The test parts consist of multiple-choice questions with dichotomous outcomes. The 3PL model is often appropriate for analyzing such items and we could verify that the 3PL model had the best fit also in these particular cases. 

\section{Methods}\label{Methods}

\subsection{Calibration of new items}\label{calnesitems}

The method we propose here for calibrating new items, can be performed under a separate offline setup; or an integrated online setup. For both setups, the estimated ability is used to determine which examinee will calibrate which item. The items used for estimating the ability are called the operational items, or the operational test. SweSAT uses a separate offline calibration, as described in Section \ref{swesat}. For an integrated setup, the calibration is usually at the end of the test since the estimated abilities have better precision when the estimation can be based on more operational items.
In this paper, we are using item parameters that are estimated from the SweSAT. We divide the items into an operational and calibration part, consisting of the same number of items. Which examinee that should calibrate which item is either determined randomly (random design) or according to the restricted optimal design defined in Section \ref{Optimaldesign}. The optimal design should theoretically produce estimates with lowest standard error of the parameters in the item parameter vector $\beta_i$ in the 3PL IRT model (Equation \ref{eq1}).

\subsection{Block Design}\label{Blockdesign}
An assumption in the \texttt{optical} exchange algorithm \citep{ULHASSAN2021107177} is that an examinee can calibrate at most one item out of a block of a low number of calibration items. We assume therefore that the calibration items are divided into $l$ blocks consisting of $m$ items each. The \texttt{optical} algorithm is then run separately on each block, assigning every examinee one item per block. Since the design is locally optimal, the calibration item parameters will need to have an initial value determined, e.g. based on a pre-estimation using a small number of examinees. 

Every examinee will be given $l$ calibration items in addition to the operational items. Which item from each block that should be assigned to the examinees is decided by the estimate of the ability of the examinees, which is estimated based on their responses to the operational test. 

Which calibration items will be assigned to which block is decided by the difficulty parameter of the items, either estimated from the full SweSAT data or pre-estimated from a smaller number of examinees. The items could be assigned to the blocks in many different ways. A main idea of design optimization is however that we aim to target the difficulty of the items with the right abilities of the examinees. Therefore, in this paper we choose to assign items to blocks such that each block should have a mix of items with different difficulty levels. We assigned the easiest item to Block 1, the second easiest to Block 2, etc. The $l$-th easiest item will be assigned to block $l$. Then the $l+1$ easiest item is assigned to Block 1, the $l+2$ easiest to Block 2 etc. This procedure is continued until all calibration items are assigned to one of the $l$ blocks. Table \ref{tab:blockdesign} shows an example of a block. This way of assigning the items to the blocks ensures a spread in difficulty between all items within a block. If there would be items with similar difficulty within a block, we would not expect large possibilities to increase efficiency with choosing an optimal design. For example, if we would use four items with $a$- and $c$-parameter like in Table \ref{tab:blockdesign} and $b$-parameter equal to -0.306 (the average of the $b$-parameters in Table \ref{tab:blockdesign}), the D-efficiency would be 1.094, i.e.\ the information gain compared to the random design would be 9.4\% instead of 12.8\%.

\begin{table}
\caption{One block of pre-estimated parameters.}
\label{tab:blockdesign}
\medskip
\begin{tabular}{lrrr} 
\toprule
& \multicolumn{1}{c}{$a$} & 
\multicolumn{1}{c}{$b$} & \multicolumn{1}{c}{$c$} \\
\midrule
1 & 0.862 & -1.063 & 0.203    \\ 
2 & 1.320 & -0.549 & 0.195    \\ 
3 & 1.220 & -0.067 & 0.155    \\ 
4 & 2.173 & 0.454 & 0.107    \\ 
\bottomrule
\end{tabular}
\end{table}

\subsection{Simulation setup -- 4 cases}\label{simulation}
The proposed method will be evaluated using simulation studies divided into four separate scenarios. An outline of the elements in the simulations under the 4 cases are illustrated in Figure \ref{fi_simulationmap}. The different cases range from a purely theoretical case to a case replicating a real calibration setting as closely as possible. There are two intermediate steps, where we relax one factor at time, aiming to isolate the influence of each. 

The simulation study compares an optimal design allocation to a random allocation of calibration items to examinees, in terms of precision of the estimated item parameters. The simulations are run $S=2000$ times. Let the estimated parameter vector for item $i, i=1,\dots,I,$ in simulation run $s, s=1,\dots,S,$ when design $d, \, d=O \, (Optimal), \, R \, (Random)$, is used be
\begin{equation}\label{eq4}
  \hat \beta^{(d)}_{i,s} = \left(
    \hat a^{(d)}_{i,s},
    \hat b^{(d)}_{i,s},
    \hat c^{(d)}_{i,s}
  \right)^\top
\end{equation}
and the true parameter vector be $\beta_i=(a_i,b_i,c_i)^\top$.

\begin{figure}[h!t]
    \centering
    \tikzstyle{datalong} = [rectangle, text width=4.5cm, minimum height=1cm,text centered, draw=black]
    \tikzstyle{data} = [rectangle, text width=3.5cm, minimum height=1cm,text centered, draw=black]
    \tikzstyle{estimation} = [rectangle, rounded corners, text width=3cm, minimum height=1cm,text centered, draw=black]
    \tikzstyle{calculation} = [trapezium, trapezium left angle=70, trapezium right angle=110, text width=2.5cm, minimum height=1cm, text centered, draw=black]
    \tikzstyle{arrow} = [thick,->,>=stealth]
    \tikzstyle{altarrow} = [thick,dashed,->,>=stealth]
    
\begin{tikzpicture}[node distance=2cm]
      \node (dataswesat) [datalong] {Data SweSAT 2018, $N=39321$ examinees, 80 items; 40 operational and 40 calibration items};
      \node (trueip) [estimation, below of=dataswesat, xshift=-6cm, yshift=-1cm] {Item parameters estimated $\hat \beta_{SweSAT} \rightarrow \beta_{true}$};
      \node (sampleswesat) [data, below of=dataswesat, xshift=4cm, yshift=-1cm] {Sample of SweSAT data, $n=200$, $I=80$};
      \node (estip) [estimation, below of=sampleswesat, xshift=0.5cm, yshift=0cm] {Item parameters $\hat \beta_{pre-est}$ estimated};
      \node (optical1) [data, below of=trueip, xshift=-0.5cm, yshift=-0.5cm] {Block formation and optimal design rules calculated with \texttt{optical} - based on $\beta_{true}$};
      \node (optical2) [data, below of=estip, xshift=1cm, yshift=-1cm] {Block formation and optimal design rules calculated with \texttt{optical} - based on $\hat \beta_{pre-est}$};
      \node (thetasim) [data, below of=dataswesat, xshift=-2cm, yshift=-1cm] {$N=39321$ abilities $\theta_{true}$ simulated from $N(0,1)$};
      \node (respmat) [data, below of=thetasim, xshift=0cm, yshift=0cm] {Response matrix generated based on $\theta_{true}$ and $\beta_{true}$};

      \node (thetaest) [estimation, below of=dataswesat, xshift=1.5cm, yshift=-6.5cm] {Abilities estimated from operational items and standardized with percentile method - $\hat \theta$};
      \node (caldes1) [data, below of=thetaest, xshift=-6.5cm, yshift=-2.5cm] {Optimal design allocation based on $\theta_{true}$ and \texttt{optical} rules using $\beta_{true}$};
      \node (caldes2) [data, below of=thetaest, xshift=-2cm, yshift=-2.5cm] {Optimal design allocation based on $\hat \theta$ and \texttt{optical} rules using $\beta_{true}$};
      \node (caldes3) [data, below of=thetaest, xshift=2.5cm, yshift=-2.5cm] {Optimal design allocation based on $\hat \theta$ and \texttt{optical} rules using $\hat \beta_{pre-est}$};
      \node (betaest1) [estimation, below of=caldes1, xshift=0cm, yshift=-0.5cm] {Item parameters estimated};
      \node (betaest2) [estimation, below of=caldes2, xshift=0cm, yshift=-0.5cm] {Item parameters estimated};
      \node (betaest3) [estimation, below of=caldes3, xshift=0cm, yshift=-0.5cm] {Item parameters estimated};
      \node (effteor) [calculation, below of=optical1, xshift=-1.5cm, yshift=-3cm] {Computation of theoretical efficiency \textit{Case I}};
      \node (eff1) [calculation, below of=betaest1, xshift=0cm, yshift=0cm] {Computation of efficiency \textit{Case II}};
      \node (eff2) [calculation, below of=betaest2, xshift=0cm, yshift=0cm] {Computation of efficiency \textit{Case III}};
      \node (eff3) [calculation, below of=betaest3, xshift=0cm, yshift=0cm] {Computation of efficiency \textit{Case IV}};

      \draw [arrow] (dataswesat) -- (sampleswesat);
      \draw [arrow] (dataswesat) -- (trueip);
      \draw [arrow] (trueip) -- (optical1);
      \draw [arrow] (dataswesat) -- (thetaest);
      \draw [arrow] (sampleswesat) -- (estip);
      \draw [arrow] (estip) -- (optical2);
      \draw [arrow] (thetasim) -- (respmat);
      \draw [arrow] (respmat) -- (caldes1);
      \draw [arrow] (optical1) -- (caldes1);
      \draw [arrow] (optical1) -- (caldes2);
      \draw [arrow] (respmat) -- (thetaest);
      \draw [arrow] (thetaest) -- (caldes2);
      \draw [arrow] (optical2) -- (caldes3);
      \draw [arrow] (thetaest) -- (caldes3);
      \draw [arrow] (caldes1) -- (betaest1);
      \draw [arrow] (caldes2) -- (betaest2);
      \draw [arrow] (caldes3) -- (betaest3);
      \draw [arrow] (optical1) -- (effteor);
      \draw [arrow] (betaest1) -- (eff1);
      \draw [arrow] (betaest2) -- (eff2);
      \draw [arrow] (betaest3) -- (eff3);   
    \end{tikzpicture}
    \caption{Outline of the simulation studies. Data and design generation is marked with rectangles, estimation of abilities or item parameters is marked as rectangles with rounded corners, and computations for result generation are marked as parallelograms.}
    \label{fi_simulationmap}
\end{figure}

We use the SweSAT data as a starting point with the aim to replicate the SweSAT test and calibration setting in the simulations. The total number of items of the quantitative part of the SweSAT is $80$, $39321$ examinees took the test and the $39321$ $\times$ $80$ response matrix is used to estimate the item parameters for the $80$ items. These are then utilized as the true parameters $\beta_i$, when generating a response matrix in each simulation iteration via the 3PL model.  Using the same number of examinees as in the SweSAT, $39321$ abilities are drawn from the $N(0,1)$ distribution. These will function as the true abilities of the examinees, $\theta_{true}$, and are also used when generating the simulated response matrices (together with the "true" item parameters $\beta_{true}$). 

We decided to let the operational and calibration part consist of $40$ items each, since each part in SweSAT consists of 40 items. The calibration part will be divided into $l=10$ blocks with $m=4$ items per block. This means that every examinee will calibrate $10$ items. The $80$ true items are randomly divided into an operational part and a calibration part, both of length $40$. Responses are simulated by generating draws from the binomial distribution $Bin(1; p_{ij}(\theta_{j}|\beta_i))$ where $p_{ij}$ are defined as in Equation \ref{eq1}. The R-package \texttt{mirt} \citep{mirt2023} was used for both the parameter estimation and generation of response matrices.

\subsubsection{\textit{Case I} - Theoretical}

The first step is to form the blocks of items, so that the optimal design allocation rules can be derived through \texttt{optical}. The blocks are constructed based on the true difficulty parameters $b_i$. The calibration items are sorted from easiest to hardest, and the blocks are created as described in \ref{Blockdesign}. The \texttt{optical} algorithm is then run for all blocks, creating the optimal design rules for each block and calculating the theoretical design efficiency. This case does not involve any data generation or parameter estimation once the true parameters $\beta_{true}$ are generated, just theoretical calculations of the efficiency of the optimal design allocation compared to a random allocation.

\subsubsection{\textit{Case II} - True abilities}

For this case, as well as for the two following cases, we generate response matrices based on $\theta_{true}$ and  $\beta_i = \beta_{true}$. The optimal design rules derived with \texttt{optical} using $\beta_{true}$ are used to assign calibration items to examinees, based on examinee ability, one item from each block. Here we use the true abilities $\theta_{true}$ also for design allocation.  Every examinee is now assigned responses for these items,  $10$ calibration items each (the non-calibration items will be set to missing in the generated response matrix) and the calibration item parameters are estimated. These item estimates are the optimal estimates $\hat \beta^{(O)}_{i,s}$.  

For the random design every examinee is randomly assigned responses for $10$ calibration items (and missing values for the rest of the items). The item parameters for the calibration items are estimated using these $10$ responses for every respondent, yielding the random design estimates of the item parameters $\hat \beta^{(R)}_{i,s}$.

Design efficiencies are then calculated using the item parameter estimates of the two designs over the $S=2000$ simulation runs. 

\subsubsection{\textit{Case III} - Estimated abilities and true parameters}

In a real calibration situation, examinee abilities are not known and need to be estimated. We use the operational part of the response matrix to first estimate the ability of each examinee $\hat \theta_i^{(0)}$; for details see Section \ref{sec:thetahat0}. The optimal design rules derived with \texttt{optical} using $\beta_{true}$ are then again used to assign calibration items to examinees, but this time based on the estimated examinee ability. The efficiencies of the resulting item parameter estimates compared to the random allocation are calculated over the $S=2000$ simulation runs.

\subsubsection{\textit{Case IV} - Estimated abilities and pre-estimated parameters}

To come even closer to a realistic setup, we relax also the fact that the item parameters used in the design generation are known. Instead, we simulate $200$ responses and pre-estimate values for the calibration items $\hat \beta_{pre-est}$. We use Bayesian estimation (through R-stan) to avoid the severe bias that otherwise can appear when using ML-estimation for a small number of examinees. The reason for only letting $200$ responses pre-estimate the calibration items is that, in a real test situation, you would like to keep the costs down. The pre-estimates are then used for block formation and derivation of the optimal allocation rules through \texttt{optical}. 
The optical allocation rules that dictate which items the examinees will calibrate are now dependent on their estimated abilities $\hat \theta_i^{(0)}$. Item parameter estimates from the optimal design allocation under this setup are set in relation to the parameter estimates from the random design, again averaged over the $S=2000$ simulation runs.

\subsubsection{Estimation of parameters}

In all cases where calibration item parameters are estimated, the abilities estimated from the operational test will be treated as fixed during subsequent estimation. This is the so-called Method A \citep{Stocking1988}, or fixed ability item parameter calibration. If the ability is not treated as fix, the EM-algorithm used for estimating the abilities will assume that the examinees abilities follow a standard normal distribution. Since the examinees calibrating a certain item are purposely selected when the optimal design allocation is used, their abilities do not follow the assumed distribution, and the estimates would be biased.

\cite{ban2001comparative} evaluated Method A among other different estimation methods in the setting of online CAT and concluded that the Multiple Expectation-Maximisation (MEM) performed the best. Method A was associated with the highest error and the treating of abilities as fixed was pointed out as a theoretical weakness. Nevertheless, for our purposes the possibility to set the abilities according to the distribution used to derive the optimal allocation is essential. Since Method A ignores the estimation error of the ability estimates, the estimates of the calibration items will be inaccurate to some extent. This problem is more severe with shorter operational tests than we suggests. An alternative to the Method A, especially with shorter operational test length, could be the method proposed in \cite{HeChenLi2017}. 

\subsubsection{Estimation of abilities}
\label{sec:thetahat0}

\begin{figure}[th]
    \centering
    \includegraphics[width=80mm]{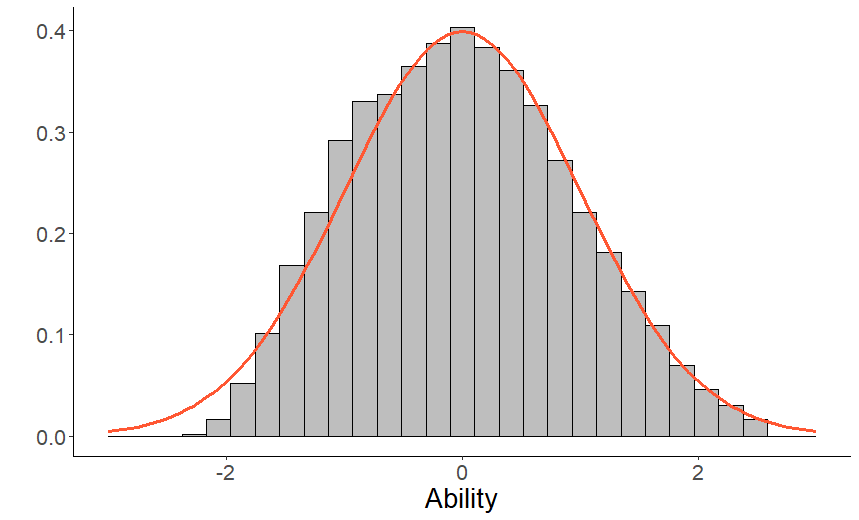}
    \caption{Histogram of estimated abilities $\hat \theta_i^{(0)}$ in one simulation run and standard normal density as comparison.}
    \label{fig:hist_est_abilities}
\end{figure}

In \textit{Case III} and \textit{Case IV} the abilities of the examinees are estimated. Based on the results from the operational part of the test, the ability of each examinee is estimated using the Expected A Posteriori (EAP) method \citep{Baker2004}. Unfortunately, despite that the true abilities are known to be $N(0,1)$ in our simulation studies, the EAP-estimates have a non-normal distribution. The negative abilities are somewhat shrunken towards 0, see Figure \ref{fig:hist_est_abilities}, which is due to the asymmetry of the 3PL model. 

Since the optimal design method assumes a standard normally distributed population, we transform the abilities $\hat \theta_i^{(0)}$ with the percentile method to values $\hat\theta_i$ distributed as $N(0,1)$, see e.g.\ \cite{Kolen2014}, Section 9.5.2.
The percentile-transformed estimates $\hat\theta_i$ are then used to determine the calibration item allocation for the optimal design.

\subsection{Definitions of measures}\label{Defofmeasures}

In the following section, several summary measures that are used to evaluate the designs in the simulation studies are presented. For the sake of clarity in the presentation, the index $i$ will be temporarily dropped from the formulas in the following section. Note that each measure is defined per item, although the index $i$ is not explicitly written out.

\subsubsection{Error matrix}\label{errormatrix}

The {\bf empirical error matrix} can be viewed as a multivariate version of the mean squared error of an estimator. When design $d, d=O, R,$ is used, it is computed as
\begin{equation}\label{eq5}
   \mathrm{Q}(\hat{\beta}^{(d)};\beta) =  
   \frac{1}{S}\sum_{s=1}^S (\hat\beta^{(d)}_{s}-\beta) (\hat\beta^{(d)}_{s}-\beta)^\top.
\end{equation}
 Based on this, the {\bf empirical  D-criterion} is defined as
\begin{equation}\label{eq6}
  \mathrm{D}(\hat{\beta}^{(d)};\beta)= \det\hspace{2pt}[ \mathrm{Q}(\hat{\beta}^{(d)};\beta)].
\end{equation}


\subsubsection{MSE}\label{MSE}

The {\bf empirical mean squared error} for the parameter estimators $\hat{\beta}$ in relation to the true parameters $\beta$ is defined by the $3\times 1$ vector
\begin{equation}\label{eq7}
  \mathrm{MSE}(\hat{\beta}^{(d)};\beta) = 
  \frac{1}{S}\sum_{s=1}^S \left((\hat a^{(d)}_{s}-a)^2,(b^{(d)}_{s}-b)^2,(c^{(d)}_{s}-c)^2)\right)^T
\end{equation}
This is equivalent to the diagonal of the empirical error matrix defined above, that is 
\begin{equation}\label{eq8}
  \mathrm{MSE}(\hat{\beta}^{(d)};\beta) = 
  \mathrm{diag}\hspace{2pt}[\mathrm{Q}(\hat{\beta}^{(d)};\beta)].
\end{equation}
To summarize the MSE of all three parameters in the model, the average MSE is computed:
\begin{equation}\label{eq9}
  \mathrm{AMSE}(\hat{\beta}^{(d)};\beta) = \left\{\mathrm{MSE}(\hat{a}^{(d)};a)+ 
  \mathrm{MSE}(\hat{b}^{(d)};b) +
  \mathrm{MSE}(\hat{c}^{(d)};c)\right\}/3.
\end{equation}
This measure is the empirical counterpart of the A-optimality criterion.

\subsubsection{CC method}\label{CC-method}

Inspired by the Haebara approach for IRT test equating \citep{Kolen2014}, the squared difference between \textit{ICC}:s based on estimated and true parameters, is evaluated for a certain ability level $\theta_j$. The \textbf{empirical characteristic curve difference} is
\begin{equation}\label{eq10}
  d(\theta_j|\hat{\beta}^{(d)};\beta) = 
  \frac{1}{S}\sum_{s=1}^S
  \left\{p\left(\theta_j|\hat{a}^{(d)}_{s},\hat{b}^{(d)}_{s},\hat{c}^{(d)}_{s}\right)-p\left(\theta_j|a,b,c\right)\right\}^2,
\end{equation}
and summing over the abilities of the $j=1,\dots,N$ examinees yields the total difference 
\begin{equation}\label{eq11}
  \mathrm{CC}(\hat{\beta}^{(d)};\beta) = 
  \sum_{j=1}^N d(\theta_j|\hat{\beta}^{(d)};\beta).
\end{equation}

\bigskip

\subsection{Evaluation metrics}\label{evmatrix}

To be able to evaluate which of the two designs: \textit{Optimal (O)} or \textit{Random (R)} performs better, several metrics are calculated as ratios of the previously defined measures. 

\bigskip

\subsubsection{Relative average MSE}\label{REMSE}

The relative average MSE, or relative A-efficiency, is given by the ratio of the empirical AMSE of the random design $(R)$ and the optimal design $(O)$
\begin{equation}\label{eq12}
  \mathrm{RE}_{\mathrm{A}}=\frac{\mathrm{AMSE}(\hat{\beta}^{(R)};\beta)}{\mathrm{AMSE}(\hat{\beta}^{(O)};\beta)}.
\end{equation}

When the ratio exceeds 1, the random design is estimated less precisely and the optimal design is preferred.

\subsubsection{Relative D-efficiency}\label{RED}

The relative D-efficiency is obtained as the ratio of the empirical D-criterion of the random design $(R)$ and the optimal design $(O)$
\begin{equation}\label{eq13}
  \mathrm{RE}_{\mathrm{D}}=\left(\frac{\mathrm{D}(\hat{\beta}^{(R)};\beta)}{\mathrm{D}(\hat{\beta}^{(O)};\beta)}\right)^{\frac{1}{3}}.
\end{equation}
The exponent $1/3$ is needed since we have three parameters in the model (determinants of $3\times 3$-matrices), see e.g.\ \cite{Atkinsson}, Section 11.

When the ratio exceeds 1, the random design is associated with a higher generalized variance and the optimal design is preferred.

\subsubsection{Relative CC-efficiency}\label{RECC}

The relative CC-efficiency is defined as the ratio of the CC-criterion of the random design $(R)$ and the optimal design $(O)$
\begin{equation}\label{eq14}
  \mathrm{RE}_{\mathrm{CC}}=\frac{\mathrm{CC}(\hat{\beta}^{(R)};\beta)}{\mathrm{CC}(\hat{\beta}^{(O)};\beta)}.
\end{equation}

When the ratio exceeds 1, the random design yields a larger difference between the characteristic curves, and the optimal design is preferred.

\subsubsection{Overall evaluation}\label{overall}

All of the previously defined measures are calculated \textit{per item}. For an overall assessment across all 40 items, we take the geometric mean over them. Since efficiencies are relative measures, the geometric mean is more appropriate than the arithmetic mean. E.g.\ if Item 1 has efficiency 0.5, Item 2 has 2, then the random design needs half of the examinees for Item 1 and double the examinees for Item 2 to obtain similar precision as the optimal design. Overall, both designs should then be of the same quality which is reflected by the geometric mean of 1. In contrast, the arithmetic mean would be 1.25 suggesting an advantage of the optimal design. The arithmetic mean would give a too optimistic overall measure; the geometric mean is always smaller than or equal to the arithmetic mean. 

 
\section{Results}\label{results}
The results are presented as the evaluation metric per item based on averages taken over the $S = 2000$ simulation runs. The number of runs was chosen to obtain a precision of relative item efficiencies of $\pm 0.05$ (we have derived bootstrap 95\%-confidence interval for the simulation error in some cases which are approximately of that size but we have not included them in the results later for sake of clarity). The average relative efficiency for all calibration items has then a precision of less than $\pm 0.01$. 

All 4 simulation cases have been evaluated and the results of the overall evaluations are given in Table \ref{tab:overall}. Results per item of \textit{Case III} and \textit{Case IV} are presented in more detail in this section. The theoretical efficiences per item of \textit{Case I} are given in Table \ref{tab:theoeff} in the appendix. The results of \textit{Case II} are very similar to those, and the per item details are therefore omitted. 

Since the D-optimality criterion is the criterion that was used for optimization of the design, the D-efficiency $\mathrm{RE_D}$ will be in focus when examining the results. Plots to display the correlation between the true item parameters and the relative D-efficiency $\mathrm{RE_D}$ are given.
Corresponding plots of $\mathrm{RE_{CC}}$ and $\mathrm{RE_{A}}$ can be found in the appendix.

A general result is that the proposed method using the block design and the \texttt{optical} algorithm manages to produce item parameters estimated with better precision in a majority of the cases.

In Tables \ref{tab:evmetricsest} and \ref{tab:evmetricsnoest} it can be seen that a majority of the items have a value of $\mathrm{RE_D} >1$. Also, as expected, the three measures $\mathrm{RE_D}$, $\mathrm{RE_{CC}}$, and $\mathrm{RE}_{\mathrm{A}}$ are correlated; e.g.\ the Pearson correlation between $\mathrm{RE_D}$ and $\mathrm{RE_{CC}}$ is 0.68. When the calibration item parameters are pre-estimated the overall measure of $\mathrm{RE_D}$ are lower compared to when the item-parameters used are the true ones, as shown in Table \ref{tab:overall}. This gives us an indication of the magnitude of the efficiency loss due to the uncertainty of the estimation of item parameters. Roughly half of the overall efficency gain over the random allocation is lost due to the pre-estimation of item parameters.

\begin{figure}[H]

\centering
\includegraphics[width=0.75\linewidth]{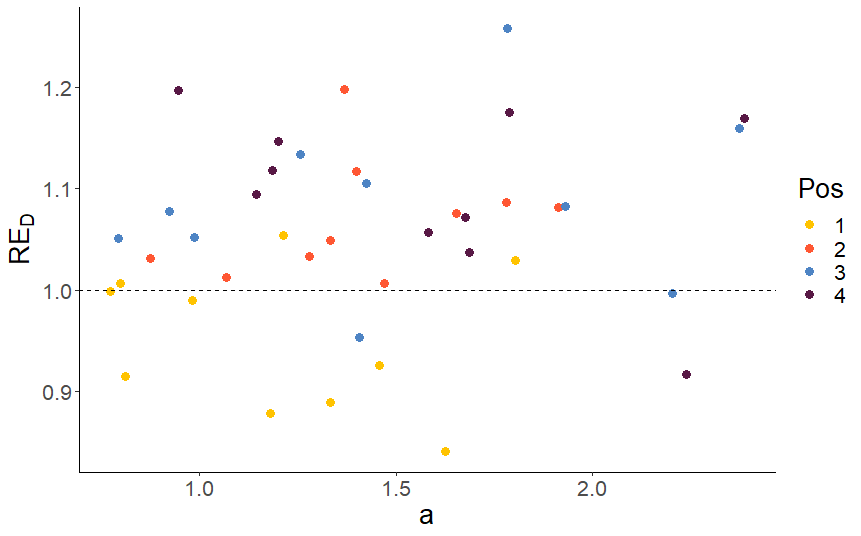}
\includegraphics[width=0.75\linewidth]{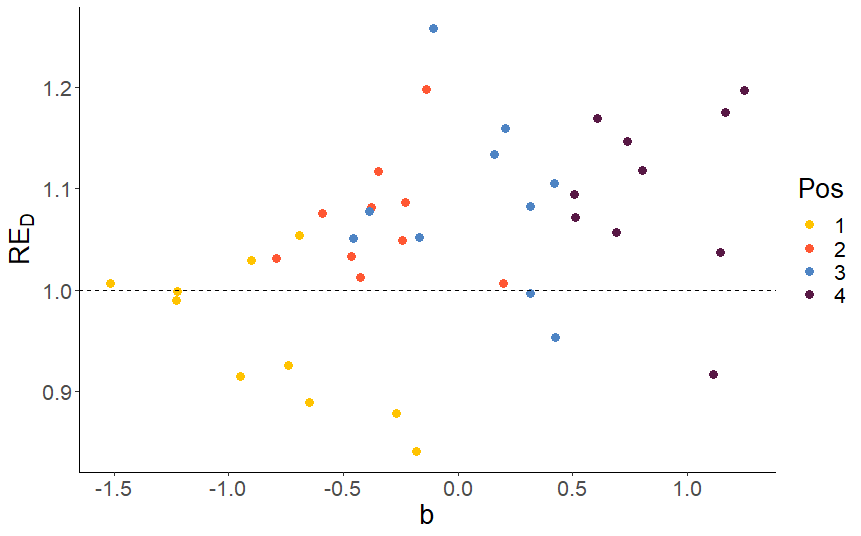}
\includegraphics[width=0.75\linewidth]{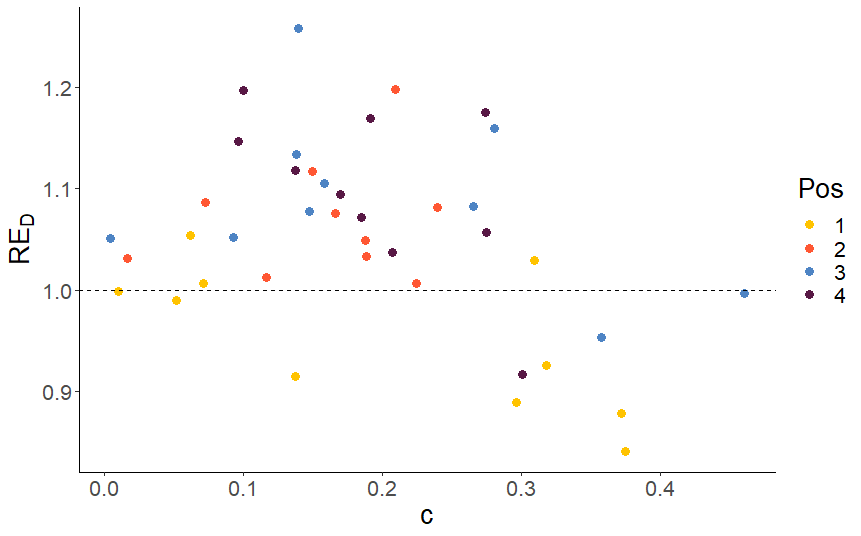}
\medskip
\caption{True item parameters plotted against $\mathrm{RE}_D$, grouped on position in the block.  \textit{Case IV}: Optimal allocation based on estimated abilities and pre-estimated parameters.}
\label{fig:estabc}
\end{figure}

\begin{figure}[H]

\centering
\includegraphics[width=0.75\linewidth]{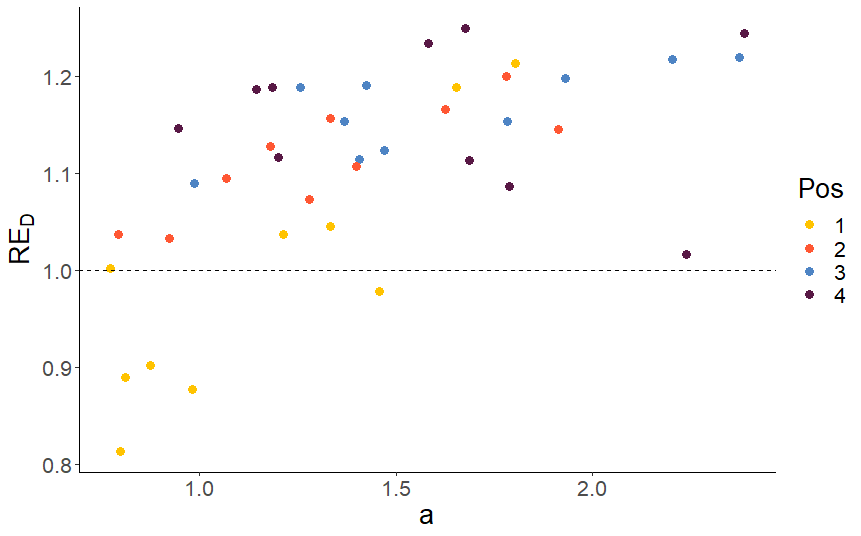}
\includegraphics[width=0.75\linewidth]{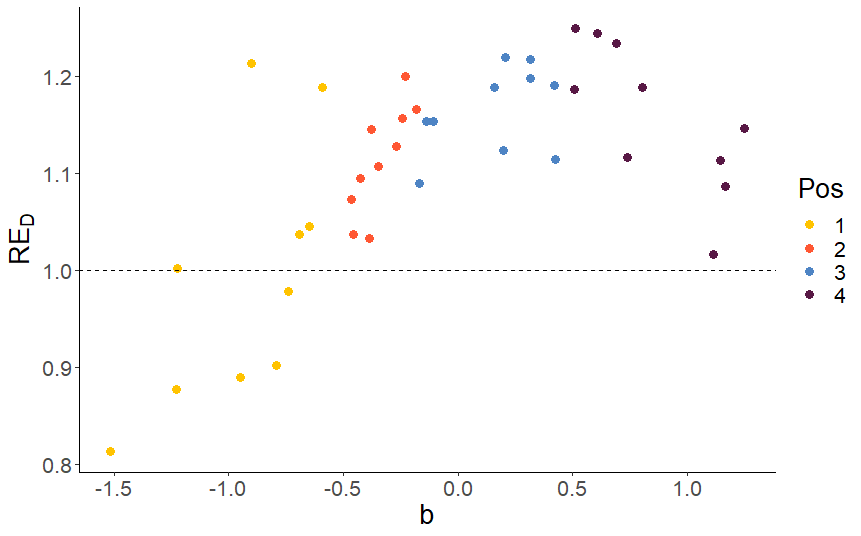}
\includegraphics[width=0.75\linewidth]{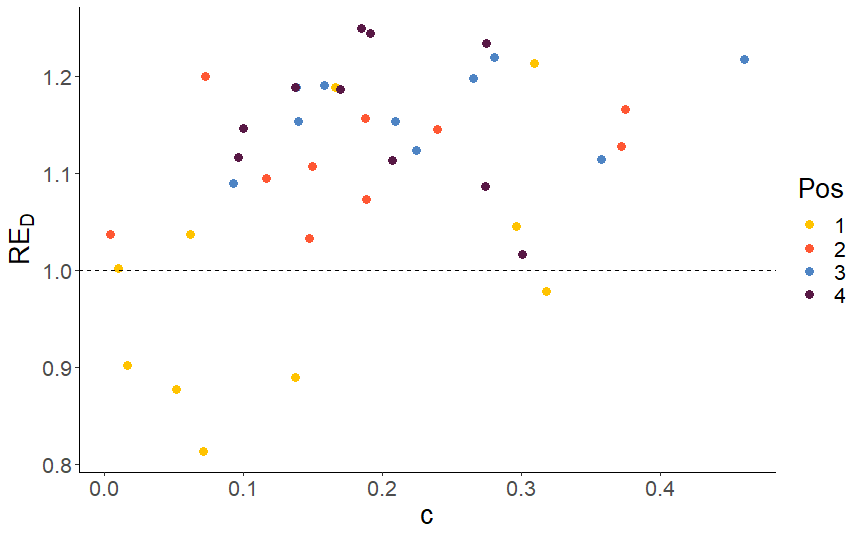}
\medskip
\caption{True item parameters plotted against $\mathrm{RE}_D$, grouped on position in the block. \textit{Case III}: optimal design allocation based on estimated abilities and true item parameters.}
\label{fig:noestabc}
\end{figure}

Figures \ref{fig:estabc} and \ref{fig:noestabc} show scatter plots of $\mathrm{RE}_\mathrm{D}$ versus true item parameters for \textit{Case IV} with pre-estimated item parameters and \textit{Case III} using true item parameters, respectively. The colors of the dots indicate the position in the blocks. The position in the block is determined by the difficulty of the item ($b_i$-parameter), which means that position is a measure of the relative difficulty in the blocks. In both cases, irrespective of which parameter we examine, it is almost always the easiest items in each block that stands out as being less precisely estimated in terms of D-efficiency $\mathrm{RE_D}$ (with just a few exceptions). The per item efficiencies generally have higher values for \textit{Case III} compared to \textit{Case IV}, since we do not have any item parameter uncertainty when the item parameters are assumed known in \textit{Case III}. A similar pattern is observed also with respect to the other two measures, as can be seen in Figures \ref{fig:estabc_CC} to \ref{fig:noestabc_A} in the Appendix.

\bigskip

\begin{table}[H]
\centering
\caption{\label{tab:overall}Relative efficiencies $\mathrm{RE_{D}}$, $\mathrm{RE_{CC}}$, $\mathrm{RE_{A}}$ for optimal versus random design, summarized for all of the 40 items for the three measures (geometric mean).}
\medskip
\begin{tabular}{lrrr} 
\toprule
 & \multicolumn{1}{c}{$\mathrm{RE_D}$} 
 & \multicolumn{1}{c}{$\mathrm{RE_{CC}}$}   
 & \multicolumn{1}{c}{$\mathrm{RE_{A}}$} \\
\midrule
\textit{Case I} : Theoretical & 1.155 & 1.126 & 1.426  \\
\textit{Case II}: True abilities & 1.155  & 1.105 & 1.346  \\ 
\textit{Case III}: Estimated abilities and true parameters & 1.103 & 1.080 & 1.198 \\ 
\textit{Case IV}: Estimated abilities and pre-estimated parameters & 1.048 & 1.048 & 1.093  \\ 
\bottomrule
\end{tabular}
\end{table}

\pagebreak
\renewcommand{\arraystretch}{0.9}

\begin{table}[H]
\caption{\label{tab:evmetricsest}Relative efficiencies $\mathrm{RE_{D}}$, $\mathrm{RE_{CC}}$, $\mathrm{RE_{A}}$ for optimal versus random design and true item parameters $a, b, c$; per item. \textit{Case IV}: Optimal allocation based on estimated abilities and pre-estimated parameters.}
\centering
\scalebox{0.95}{
\begin{tabular}{llrrrrrrr}
\toprule
\multicolumn{1}{c}{Block} & \multicolumn{1}{c}{Pos} & 
\multicolumn{1}{c}{$\mathrm{RE}_\mathrm{D}$} & \multicolumn{1}{c}{$\mathrm{RE}_{\mathrm{CC}}$}  & 
\multicolumn{1}{c}{$\mathrm{RE}_{\mathrm{A}}$} &  
\multicolumn{1}{c}{$a$} &  \multicolumn{1}{c}{$b$} &  \multicolumn{1}{c}{$c$} & \multicolumn{1}{c}{Item}\\

\midrule
1 & 1 & 1.007 & 1.026 & 1.279 & 0.799 & -1.516 & 0.071 & 39 \\ 
1 & 2 & 1.033 & 1.090 & 1.079 & 1.281 & -0.466 & 0.189 & 8 \\ 
1 & 3 & 1.077 & 1.035 & 1.255 & 0.924 & -0.389 & 0.147 & 17 \\  \medskip
1 & 4 & 1.169 & 1.119 & 1.135 & 2.385 & 0.607 & 0.191 & 25 \\

2 & 1 & 1.029 & 1.344 & 1.254 & 1.803 & -0.900 & 0.310 & 20 \\ 
2 & 2 & 1.081 & 1.136 & 1.138 & 1.911 & -0.376 & 0.240 & 11 \\ 
2 & 3 & 0.997 & 0.951 & 0.850 & 2.203 & 0.315 & 0.461 & 26 \\  \medskip
2 & 4 & 1.056 & 1.016 & 1.023 & 1.582 & 0.694 & 0.275 & 19 \\ 

3 & 1 & 0.925 & 1.041 & 0.861 & 1.458 & -0.739 & 0.318 & 3 \\ 
3 & 2 & 1.117 & 1.103 & 1.330 & 1.399 & -0.346 & 0.150 & 35 \\ 
3 & 3 & 1.258 & 1.174 & 1.337 & 1.782 & -0.108 & 0.140 & 2 \\  \medskip
3 & 4 & 1.072 & 1.003 & 1.126 & 1.676 & 0.513 & 0.185 & 29\\ 

4 & 1 & 0.889 & 0.922 & 0.782 & 1.333 & -0.648 & 0.296 & 21 \\ 
4 & 2 & 1.031 & 1.037 & 1.161 & 0.876 & -0.792 & 0.016 & 27 \\ 
4 & 3 & 1.052 & 1.036 & 0.999 & 0.989 & -0.170 & 0.093 & 38 \\  \medskip
4 & 4 & 1.094 & 1.032 & 1.238 & 1.144 & 0.510 & 0.169 & 40 \\ 

5 & 1 & 0.990 & 1.037 & 0.954 & 0.983 & -1.227 & 0.052 & 31 \\ 
5 & 2 & 1.049 & 1.050 & 1.152 & 1.334 & -0.241 & 0.187 & 4 \\ 
5 & 3 & 1.083 & 1.099 & 1.136 & 1.930 & 0.315 & 0.266 & 24 \\  \medskip
5 & 4 & 1.118 & 1.017 & 1.195 & 1.187 & 0.806 & 0.137 & 18 \\ 

6 & 1 & 0.841 & 0.898 & 0.735 & 1.627 & -0.184 & 0.375 & 1 \\ 
6 & 2 & 1.087 & 1.075 & 1.238 & 1.781 & -0.228 & 0.073 & 5 \\ 
6 & 3 & 1.105 & 1.077 & 1.157 & 1.426 & 0.421 & 0.158 & 12 \\  \medskip
6 & 4 & 0.917 & 1.010 & 0.832 & 2.238 & 1.116 & 0.301 & 6 \\ 

7 & 1 & 0.999 & 1.006 & 1.314 & 0.775 & -1.223 & 0.010 & 36 \\ 
7 & 2 & 1.013 & 1.057 & 1.052 & 1.070 & -0.428 & 0.116 & 37 \\ 
7 & 3 & 1.160 & 1.130 & 1.216 & 2.374 & 0.208 & 0.281 & 23 \\  \medskip
7 & 4 & 1.175 & 1.138 & 1.153 & 1.789 & 1.169 & 0.274 & 30\\ 

8 & 1 & 0.878 & 0.864 & 0.793 & 1.181 & -0.270 & 0.372 & 7 \\ 
8 & 2 & 1.075 & 0.975 & 0.992 & 1.652 & -0.590 & 0.166 & 22 \\ 
8 & 3 & 1.050 & 1.035 & 1.281 & 0.794 & -0.454 & 0.004 & 14 \\  \medskip
8 & 4 & 1.197 & 1.136 & 1.207 & 0.946 & 1.250 & 0.100 & 13 \\ 

9 & 1 & 0.915 & 0.919 & 1.000 & 0.813 & -0.949 & 0.137 & 15 \\ 
9 & 2 & 1.198 & 1.155 & 1.242 & 1.369 & -0.140 & 0.210 & 33\\ 
9 & 3 & 0.953 & 0.935 & 1.193 & 1.407 & 0.427 & 0.358 & 16 \\  \medskip
9 & 4 & 1.147 & 1.114 & 1.116 & 1.200 & 0.741 & 0.096 & 34 \\ 

10 & 1 & 1.054 & 1.136 & 1.237 & 1.214 & -0.691 & 0.062 & 32 \\ 
10 & 2 & 1.006 & 1.030 & 0.968 & 1.471 & 0.196 & 0.225 & 28 \\ 
10 & 3 & 1.134 & 1.074 & 1.245 & 1.256 & 0.157 & 0.138 & 9 \\  
10 & 4 & 1.037 & 1.026 & 0.998 & 1.687 & 1.147 & 0.207 & 10 \\ 
\bottomrule
\end{tabular}}
\end{table}

\pagebreak
\renewcommand{\arraystretch}{0.9}
\begin{small}
\begin{table}[H]
\caption{\label{tab:evmetricsnoest}Relative efficiencies $\mathrm{RE_{D}}$, $\mathrm{RE_{CC}}$, $\mathrm{RE_{A}}$ for optimal versus random design and true item parameters $a, b, c$; per item. \textit{Case III}: optimal design allocation based on estimated abilities and true item parameters.}
\centering
\scalebox{0.95}{
\begin{tabular}{llrrrrrrr}
\toprule
\multicolumn{1}{c}{Block} & \multicolumn{1}{c}{Pos} & 
\multicolumn{1}{c}{$\mathrm{RE}_\mathrm{D}$} & \multicolumn{1}{c}{$\mathrm{RE}_{\mathrm{CC}}$}  & 
\multicolumn{1}{c}{$\mathrm{RE}_{\mathrm{A}}$} &  
\multicolumn{1}{c}{$a$} &  \multicolumn{1}{c}{$b$} &  \multicolumn{1}{c}{$c$} & \multicolumn{1}{c}{Item}\\
\midrule
1 & 1 & 0.814 & 0.808 & 0.693 & 0.799 & -1.516 & 0.071 & 39 \\ 
1 & 2 & 1.073 & 1.079 & 1.017 & 1.281 & -0.466 & 0.189 & 8 \\ 
1 & 3 & 1.090 & 1.028 & 1.245 & 0.989 & -0.170 & 0.093 & 38 \\ \medskip
1 & 4 & 1.187 & 1.081 & 1.443 & 1.144 & 0.510 & 0.169 & 40 \\ 
2 & 1 & 0.878 & 0.857 & 0.768 & 0.983 & -1.227 & 0.052 & 31 \\ 
2 & 2 & 1.038 & 0.931 & 1.511 & 0.794 & -0.454 & 0.004 & 14 \\ 
2 & 3 & 1.154 & 1.119 & 1.243 & 1.369 & -0.140 & 0.210 & 33 \\ \medskip
2 & 4 & 1.250 & 1.162 & 1.295 & 1.676 & 0.513 & 0.185 & 29 \\ 
3 & 1 & 1.003 & 0.995 & 1.463 & 0.775 & -1.223 & 0.010 & 36\\ 
3 & 2 & 1.095 & 1.115 & 1.361 & 1.070 & -0.428 & 0.116 & 37\\ 
3 & 3 & 1.154 & 1.101 & 1.224 & 1.782 & -0.108 & 0.140 & 2 \\ \medskip
3 & 4 & 1.244 & 1.160 & 1.255 & 2.385 & 0.607 & 0.191 & 25 \\ 
4 & 1 & 0.890 & 0.918 & 0.832 & 0.813 & -0.949 & 0.137 & 15 \\ 
4 & 2 & 1.034 & 1.013 & 1.277 & 0.924 & -0.389 & 0.147 & 17 \\ 
4 & 3 & 1.189 & 1.153 & 1.340 & 1.256 & 0.157 & 0.138 & 9 \\ \medskip
4 & 4 & 1.235 & 1.167 & 1.395 & 1.582 & 0.694 & 0.275 & 19 \\ 
5 & 1 & 1.214 & 1.408 & 1.468 & 1.803 & -0.900 & 0.310 & 20 \\ 
5 & 2 & 1.146 & 1.149 & 1.227 & 1.911 & -0.376 & 0.240 & 11 \\
5 & 3 & 1.124 & 1.035 & 1.102 & 1.471 & 0.196 & 0.225 & 28 \\ \medskip
5 & 4 & 1.116 & 1.044 & 1.149 & 1.200 & 0.741 & 0.096 & 34 \\ 
6 & 1 & 0.902 & 0.899 & 1.055 & 0.876 & -0.792 & 0.016 & 27 \\
6 & 2 & 1.108 & 1.123 & 1.266 & 1.399 & -0.346 & 0.150 & 35 \\ 
6 & 3 & 1.219 & 1.107 & 1.210 & 2.374 & 0.208 & 0.281 & 23 \\ \medskip
6 & 4 & 1.189 & 1.070 & 1.363 & 1.187 & 0.806 & 0.137 & 18 \\ 
7 & 1 & 0.979 & 1.148 & 0.955 & 1.458 & -0.739 & 0.318 & 3 \\ 
7 & 2 & 1.128 & 1.136 & 1.479 & 1.181 & -0.270 & 0.372 & 7 \\ 
7 & 3 & 1.218 & 1.162 & 1.223 & 2.203 & 0.315 & 0.461 & 26 \\ \medskip
7 & 4 & 1.017 & 0.918 & 0.951 & 2.238 & 1.116 & 0.301 & 6 \\ 
8 & 1 & 1.037 & 1.135 & 1.148 & 1.214 & -0.691 & 0.062 & 32 \\ 
8 & 2 & 1.157 & 1.145 & 1.415 & 1.334 & -0.241 & 0.187 & 4 \\ 
8 & 3 & 1.199 & 1.166 & 1.294 & 1.930 & 0.315 & 0.266 & 24 \\ \medskip
8 & 4 & 1.114 & 1.050 & 1.015 & 1.687 & 1.147 & 0.207 & 10 \\ 
9 & 1 & 1.045 & 1.126 & 1.197 & 1.333 & -0.648 & 0.296 & 21 \\ 
9 & 2 & 1.200 & 1.177 & 1.370 & 1.781 & -0.228 & 0.073 & 5 \\ 
9 & 3 & 1.191 & 1.171 & 1.256 & 1.426 & 0.421 & 0.158 & 12 \\ \medskip
9 & 4 & 1.087 & 1.040 & 1.035 & 1.789 & 1.169 & 0.274 & 30 \\ 
10 & 1 & 1.189 & 1.327 & 1.482 & 1.652 & -0.590 & 0.166 & 22 \\ 
10 & 2 & 1.166 & 1.091 & 1.205 & 1.627 & -0.184 & 0.375 & 1 \\ 
10 & 3 & 1.115 & 1.034 & 1.169 & 1.407 & 0.427 & 0.358 & 16 \\ 
10 & 4 & 1.147 & 1.085 & 1.227 & 0.946 & 1.250 & 0.100 & 13 \\  
\bottomrule
\end{tabular}}
\end{table}
\end{small}

\section{Discussion}\label{discussion}

We studied 4 different simulation scenarios with varying degree of practicality, from entirely theoretical to the most realistic, designed to be as close as possible to the SweSAT setup. This allows us to quantify the influence of different factors (on average) by comparing the overall efficiencies, summarised over all items, between the scenarios. We conclude that there is an efficiency loss at each step, especially between cases \textit{III} and \textit{IV} (pre-estimation of item parameters) but also cases \textit{II} and \textit{III} (estimation of abilities). For the most realistic \textit{Case IV} the optimal design allocation was still better than the random design allocation, being about 5 \% more efficient in terms of RE$_D$ and RE$_{CC}$ and almost 10 \% in terms of RE$_A$.

The results per item show that the optimal design method estimates the calibration items more efficiently compared to the random design in most of the cases. Also, analyzing the block positions, we are able to identify that it is mainly for the easiest items in the blocks the optimal design method is inferior to the random allocation. This pattern can be observed both when the item parameters used in the optimization are pre-estimated (\textit{Case IV}) and when true item parameters were used (\textit{Case III}), see Figures \ref{fig:estabc} and \ref{fig:noestabc}. The reason for that is the asymmetry of the 3PL model, which implies that the $c$-parameters are estimated based on examinees with low abilities. It is more difficult to estimate $c$ when the item is easy (low $b$). Since it is more difficult to estimate $b$ and $c$ for easy items, the optimal design puts more focus on the other items in the block. Their precision can easier be improved leading to an increased overall efficiency even if accepting a decreased efficiency for the easy items. In reality, most of the examinees' abilities will however lay in $\theta \in [-2,2]$. In that span, two items can have similar \textit{Item Characteristic Curves (ICC)} even if the item parameters differ.
Since most of the abilities of examinees will lay in the span $\theta \in [-2,2]$ the consequences for using such item in an item bank might not be that severe (even if the item parameters are less precisely estimated).


As noted, the overall $\mathrm{RE}_D$ (Table \ref{tab:overall}) is lower when the calibration items are pre-estimated  (\textit{Case IV)} compared to when true item parameters are used (\textit{Case III}). For individual items there can be even larger differences, see e.g. Item 1 in Table \ref{tab:evmetricsest} and in \ref{tab:evmetricsnoest}. This indicates that the effect of putting more focus on the harder items is bigger in the scenario where the calibration items are pre-estimated. But even if the item parameters used in \texttt{optical} are not the true ones, the optimal block design seems to work for a majority of the items. When compering the two different cases it needs however to be considered that the division into the blocks is different since the pre-estimation item parameter will differ from true ones. But since the difference between the overall measures (Table \ref{tab:overall}) is not so big, it can be concluded that the method functions well when using the pre-estimates that are produced from only 200 examinees.  An alternative to pre-estimation is using experts guesses. Provided such guesses are comparable to the 200 examinee pre-estimation, the results suggest that expert guessing could also be an acceptable option.


It is shown that with the proposed method, most of the items are estimated with higher efficiency and that it is possible to identify when it is not useful. Since it is the easier items in every block that is estimated with less efficiency, it is likely that they can be identified in advance. It is reasonable to believe that the items can be ranked with respect to difficulty even though the exact value of the parameter $b$ is not known. The items that are probable to be estimated with worse efficiency for the optimal design could then be estimated through a random design for example. 

In this paper, we are not taking into account practical issues with assigning items. It could be necessary to put constraints on the calibration items, for example on items that cannot be simultaneously included in the calibration part. There must also be a mix of content in the calibration items, and they need to be compatible with the operational test items.

In the optimal design approach we are using, it is assumed that the abilities of the examinees are known. In the Cases III and IV of our study, we apply this optimal design for abilities which are estimated. This discrepancy leads to the drop in efficiency observed. However, it could be possible in future research to modify the optimal design approach to include the uncertainty around the ability estimates and to determine optimal designs for the situation of estimated abilities. With these optimal designs, it could be possible to increase efficiency in the case of estimated abilities.
Also, instead of using a pre-estimated point estimate or guess for the calibration item parameters, a prior distribution could be assigned from which an optimal-on-average (also known as Bayesian) design could be derived.

\section*{Acknowledgements}
This work was supported by the Swedish Research Council (Vetenskapsrådet), Grant 2019-02706.

\bibliography{ref}
\bibliographystyle{apalike}

\appendix
\section{}\label{appendix}

\begin{figure}[H]
\centering
\includegraphics[width=0.72\linewidth]{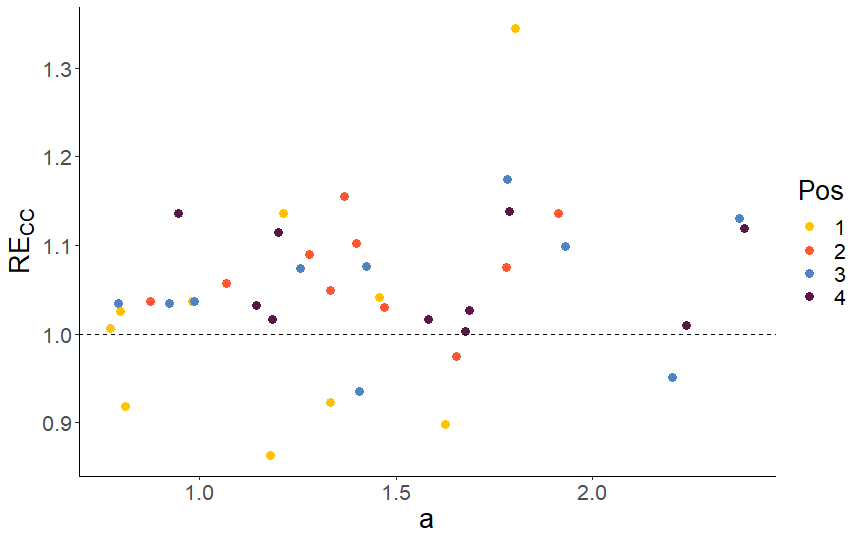}
\includegraphics[width=0.72\linewidth]{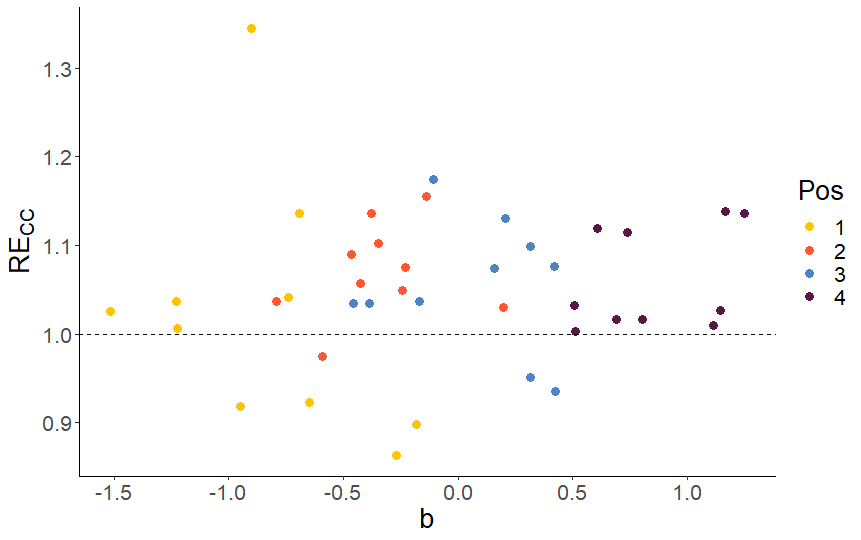}
\includegraphics[width=0.72\linewidth]{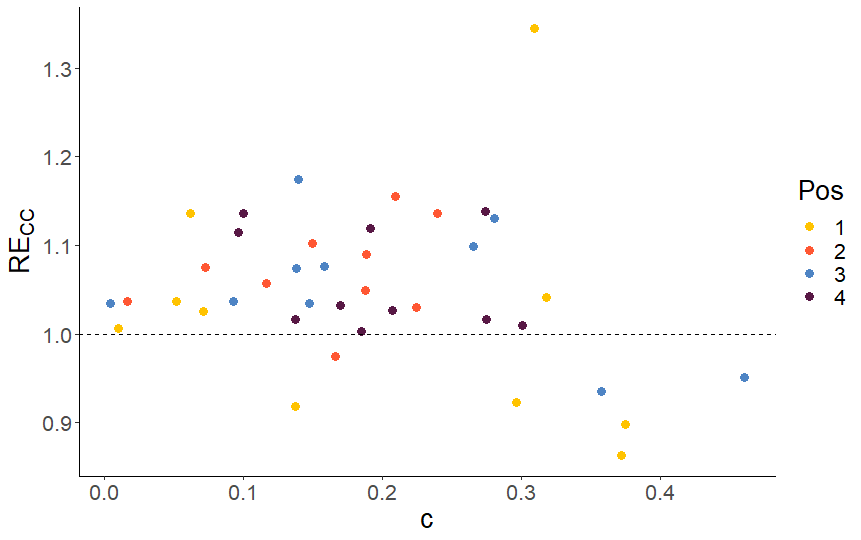}
\medskip
\caption{True item parameters plotted against $\mathrm{RE_{CC}}$, grouped on position in the block.  \textit{Case IV}: Optimal allocation based on estimated abilities and pre-estimated parameters. }
\label{fig:estabc_CC}
\end{figure}

\begin{figure}[H]
\centering
\includegraphics[width=0.72\linewidth]{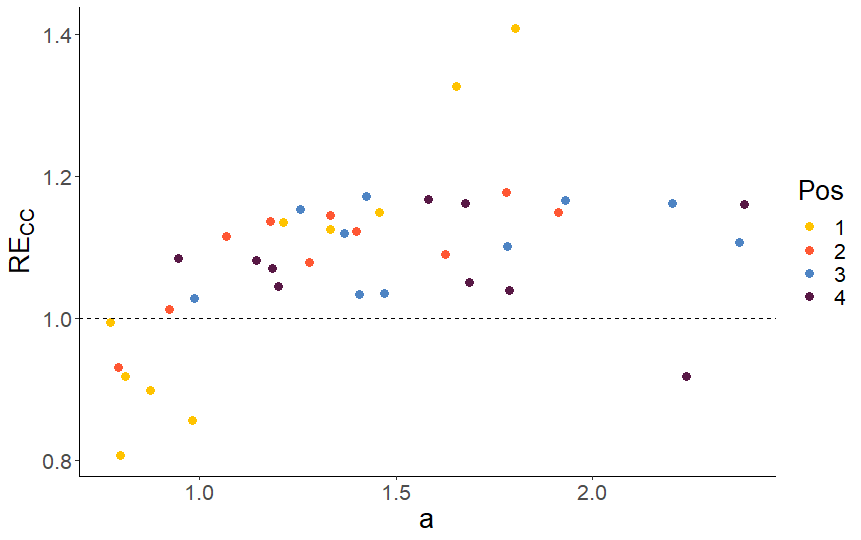}
\includegraphics[width=0.72\linewidth]{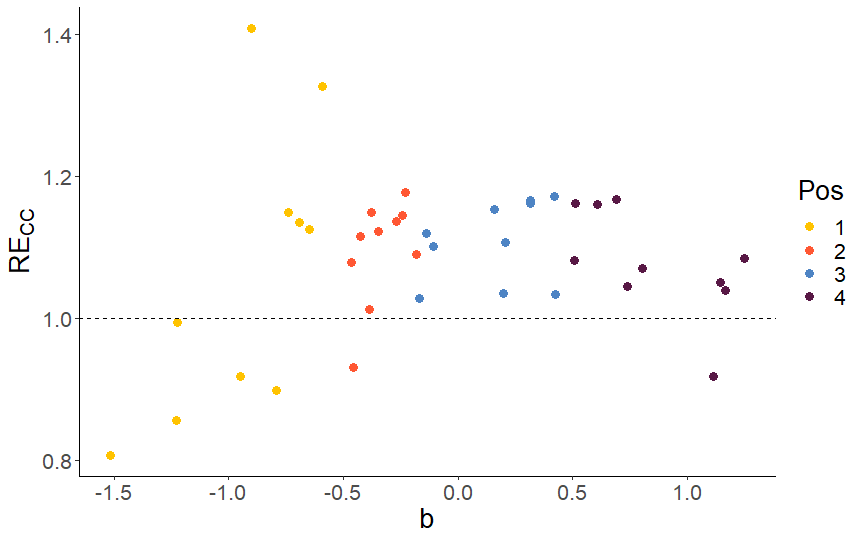}
\includegraphics[width=0.72\linewidth]{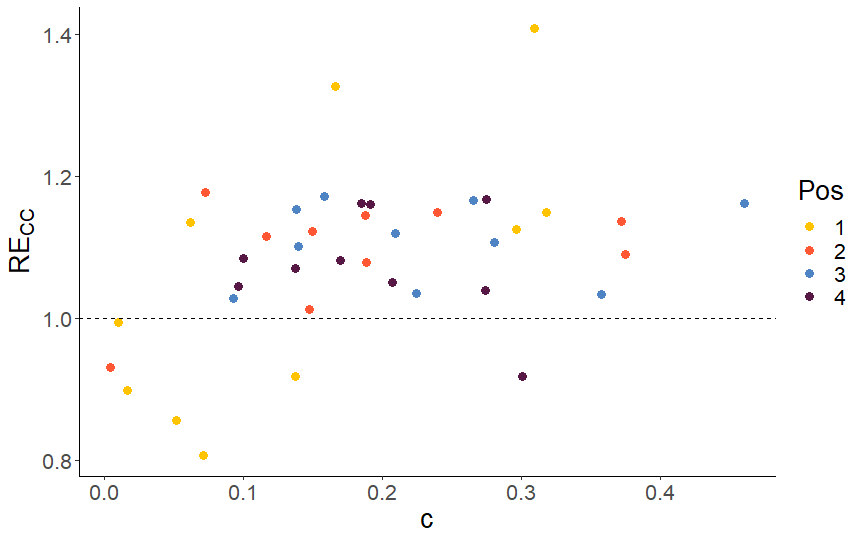}
\medskip
\caption{True item parameters plotted against $\mathrm{RE_{CC}}$, grouped on position in the block. \textit{Case III}: optimal design allocation based on estimated abilities and true item parameters.}
\label{fig:noestabc_CC}
\end{figure}

\begin{figure}[H]
\centering
\includegraphics[width=0.72\linewidth]{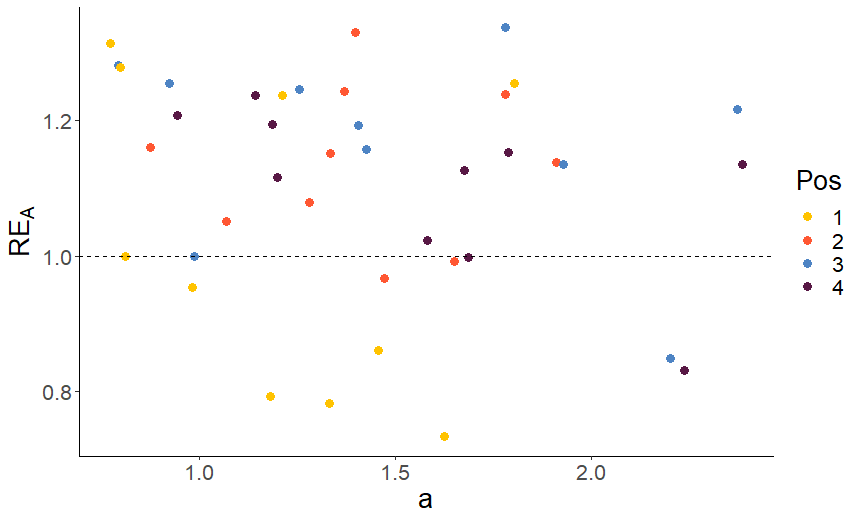}
\includegraphics[width=0.72\linewidth]{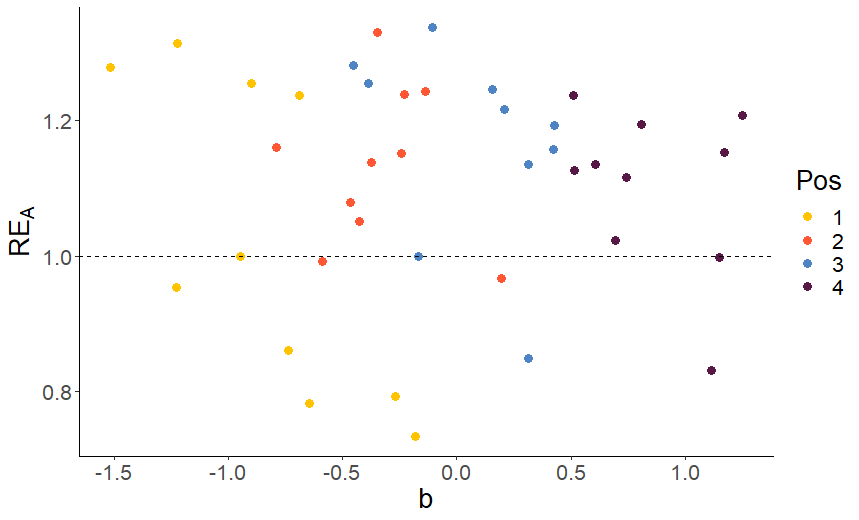}
\includegraphics[width=0.72\linewidth]{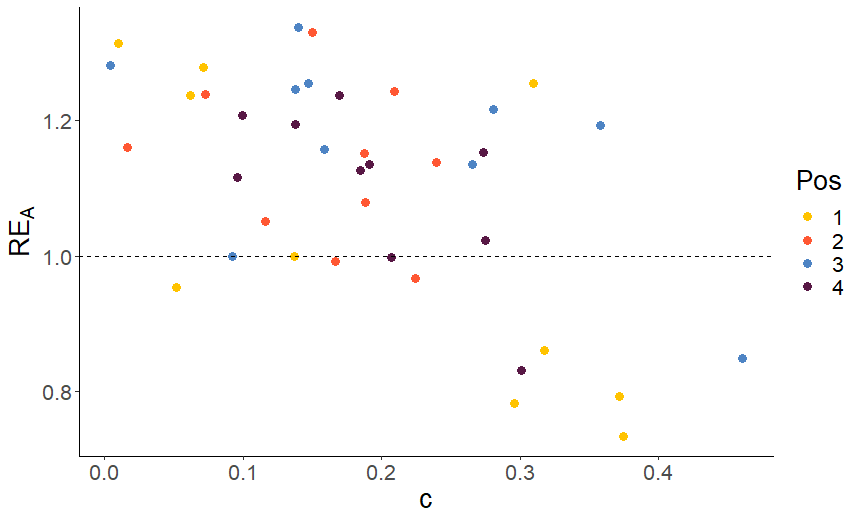}
\medskip
\caption{True item parameters plotted against $\mathrm{RE_{A}}$, grouped on position in the block.  \textit{Case IV}: Optimal allocation based on estimated abilities and pre-estimated parameters.}
\label{fig:estabc_A}
\end{figure}

\begin{figure}[H]
\centering
\includegraphics[width=0.72\linewidth]{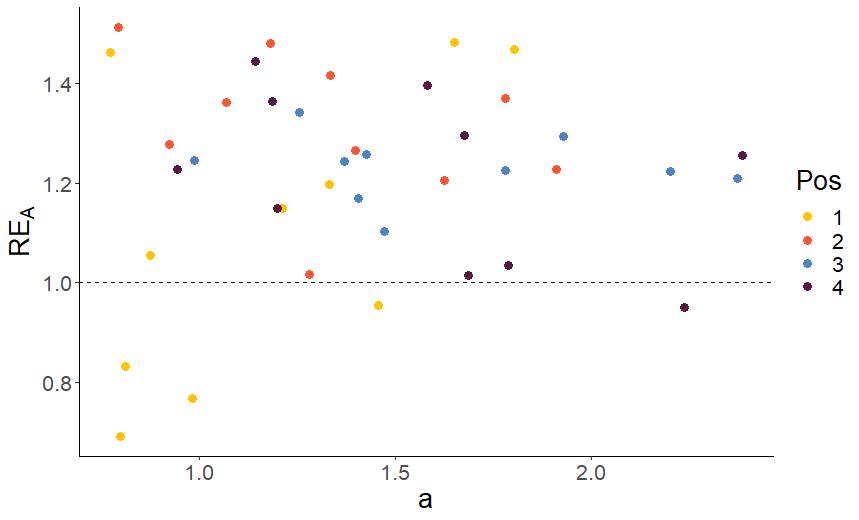}
\includegraphics[width=0.72\linewidth]{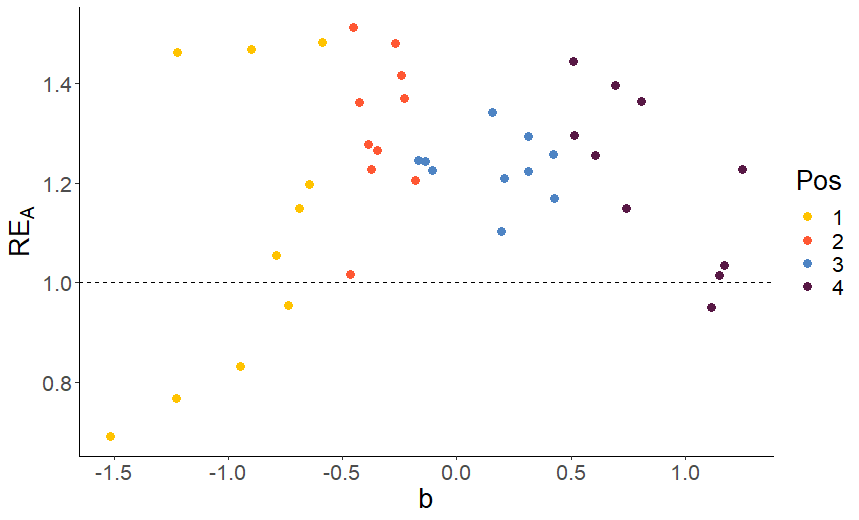}
\includegraphics[width=0.72\linewidth]{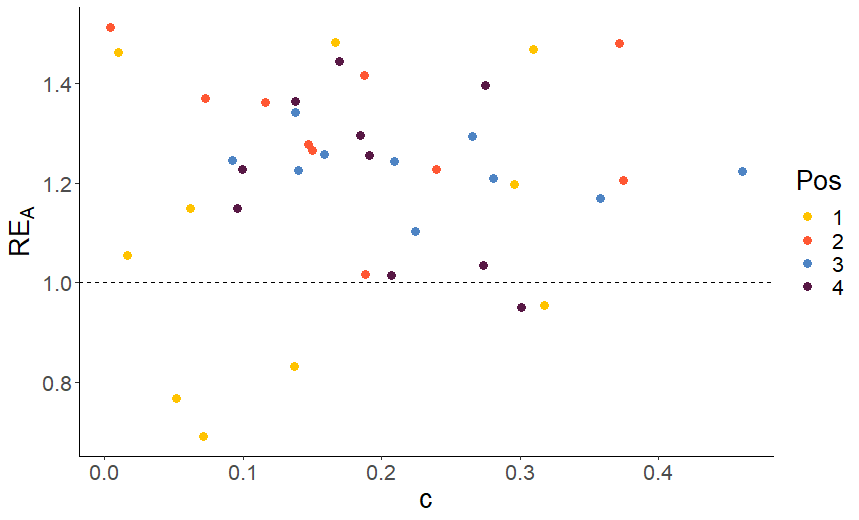}
\medskip
\caption{True item parameters plotted against $\mathrm{RE_{A}}$, grouped on position in the block. \textit{Case III}: optimal design allocation based on estimated abilities and true item parameters.}
\label{fig:noestabc_A}
\end{figure}

\begin{table}[ht]
\caption{\label{tab:theoeff}Theoreticalrelative efficiencies $\mathrm{RE_{D}}$, $\mathrm{RE_{CC}}$, $\mathrm{RE_{A}}$ for optimal versus random design and true item parameters $a, b, c$; per item. \textit{Case I}: \texttt{optical} based on true item parameters.}
\centering
\scalebox{0.95}{
\begin{tabular}{llrrrrrrr}
\toprule
\multicolumn{1}{c}{Block} & \multicolumn{1}{c}{Pos} & 
\multicolumn{1}{c}{$\mathrm{RE}_\mathrm{D}$} & \multicolumn{1}{c}{$\mathrm{RE}_{\mathrm{CC}}$}  & 
\multicolumn{1}{c}{$\mathrm{RE}_{\mathrm{A}}$} &  
\multicolumn{1}{c}{$a$} &  \multicolumn{1}{c}{$b$} &  \multicolumn{1}{c}{$c$} & \multicolumn{1}{c}{Item}\\

\midrule
1 & 1 & 1.124 & 1.043 & 1.580 & 0.799 & -1.516 & 0.071 & 39 \\ 
  1 & 2 & 1.103 & 1.083 & 1.195 & 1.281 & -0.466 & 0.189 & 8 \\ 
  1 & 3 & 1.060 & 1.070 & 1.283 & 0.989 & -0.170 & 0.093 & 38 \\ \medskip 
  1 & 4 & 1.188 & 1.199 & 1.383 & 1.144 & 0.510 & 0.169 & 40 \\ 
  
  2 & 1 & 1.136 & 1.067 & 1.612 & 0.983 & -1.227 & 0.052 & 31 \\ 
  2 & 2 & 1.076 & 1.096 & 1.474 & 0.794 & -0.454 & 0.004 & 14 \\ 
  2 & 3 & 1.146 & 1.130 & 1.252 & 1.369 & -0.140 & 0.210 & 33 \\ \medskip 
  2 & 4 & 1.226 & 1.248 & 1.453 & 1.676 & 0.513 & 0.185 & 29 \\ 
  
  3 & 1 & 1.121 & 1.052 & 1.811 & 0.775 & -1.223 & 0.010 & 36 \\ 
  3 & 2 & 1.082 & 1.056 & 1.434 & 1.070 & -0.428 & 0.116 & 37 \\ 
  3 & 3 & 1.190 & 1.138 & 1.187 & 1.782 & -0.108 & 0.140 & 2 \\ \medskip 
  3 & 4 & 1.293 & 1.329 & 1.557 & 2.385 & 0.607 & 0.191 & 25 \\ 
  
  4 & 1 & 1.076 & 1.012 & 1.494 & 0.813 & -0.949 & 0.137 & 15 \\ 
  4 & 2 & 1.052 & 1.028 & 1.350 & 0.924 & -0.389 & 0.147 & 17 \\ 
  4 & 3 & 1.105 & 1.093 & 1.208 & 1.256 & 0.157 & 0.138 & 9 \\ \medskip 
  4 & 4 & 1.224 & 1.250 & 1.499 & 1.582 & 0.694 & 0.275 & 19 \\ 
  
  5 & 1 & 1.230 & 1.100 & 1.574 & 1.803 & -0.900 & 0.310 & 20 \\ 
  5 & 2 & 1.155 & 1.196 & 1.411 & 1.911 & -0.376 & 0.240 & 11 \\ 
  5 & 3 & 1.104 & 1.176 & 1.213 & 1.471 & 0.196 & 0.225 & 28 \\ \medskip 
  5 & 4 & 1.108 & 1.050 & 1.249 & 1.200 & 0.741 & 0.096 & 34 \\ 
  
  6 & 1 & 1.149 & 1.107 & 1.866 & 0.876 & -0.792 & 0.016 & 27 \\ 
  6 & 2 & 1.143 & 1.079 & 1.314 & 1.399 & -0.346 & 0.150 & 35 \\ 
  6 & 3 & 1.252 & 1.254 & 1.325 & 2.374 & 0.208 & 0.281 & 23 \\ \medskip 
  6 & 4 & 1.191 & 1.169 & 1.503 & 1.187 & 0.806 & 0.137 & 18 \\ 
  
  7 & 1 & 1.167 & 0.974 & 1.362 & 1.458 & -0.739 & 0.318 & 3 \\ 
  7 & 2 & 1.099 & 1.025 & 1.384 & 1.181 & -0.270 & 0.372 & 7 \\ 
  7 & 3 & 1.213 & 1.218 & 1.351 & 2.203 & 0.315 & 0.461 & 26 \\ \medskip 
  7 & 4 & 1.256 & 1.363 & 1.785 & 2.238 & 1.116 & 0.301 & 6 \\ 
  
  8 & 1 & 1.157 & 1.033 & 1.619 & 1.214 & -0.691 & 0.062 & 32 \\ 
  8 & 2 & 1.125 & 1.044 & 1.365 & 1.334 & -0.241 & 0.187 & 4 \\ 
  8 & 3 & 1.187 & 1.184 & 1.408 & 1.930 & 0.315 & 0.266 & 24 \\ \medskip 
  8 & 4 & 1.261 & 1.236 & 1.855 & 1.687 & 1.147 & 0.207 & 10 \\ 
  
  9 & 1 & 1.109 & 0.997 & 1.450 & 1.333 & -0.648 & 0.296 & 21 \\ 
  9 & 2 & 1.140 & 1.089 & 1.201 & 1.781 & -0.228 & 0.073 & 5 \\ 
  9 & 3 & 1.151 & 1.165 & 1.318 & 1.426 & 0.421 & 0.158 & 12 \\ \medskip
  9 & 4 & 1.261 & 1.295 & 1.876 & 1.789 & 1.169 & 0.274 & 30 \\ 
  
  10 & 1 & 1.237 & 1.156 & 1.586 & 1.652 & -0.590 & 0.166 & 22 \\ 
  10 & 2 & 1.146 & 1.186 & 1.337 & 1.627 & -0.184 & 0.375 & 1 \\ 
  10 & 3 & 1.123 & 1.203 & 1.207 & 1.407 & 0.427 & 0.358 & 16 \\ \medskip 
  10 & 4 & 1.103 & 1.022 & 1.198 & 0.946 & 1.250 & 0.100 & 13 \\ 
  
\bottomrule
\end{tabular}}
\end{table}

\end{document}